\documentclass[a4paper,11pt]{article}
\usepackage{pos}
\usepackage{amsmath}
\title{Symmetries of temporal correlators and the nature of hot QCD}
\author*[a]{L. Ya. Glozman}
\affiliation[a]{Institute for Physics, University of Graz, Austria}
\author[b]{Y. Aoki}
\affiliation[b]{RIKEN, Kobe, Japan}
\author[c]{S. Hashimoto}
\affiliation[c]{KEK, Tsukuba, Japan}
\author[d]{C. Rohrhofer}
\affiliation[d]{Osaka University, Osaka, Japan}

\abstract{The temperature
of the chiral restoration phase transition at $\sim 130$
MeV as well as the temperature of the center symmetry ("deconfinement")
phase transition in a pure glue theory at $\sim 300$ MeV are two independent
temperatures and their interplay  determines a structure of different
regimes of hot QCD. Given a chiral spin symmetry of the color charge
and of the chromoelectric interaction we can conclude from  observed symmetries of spatial and temporal correlators of $N_F=2$ QCD with domain wall Dirac
operator at physical quark masses that above the chiral symmetry restoration
crossover around $T_{pc}$ but below rougly $3T_{pc}$ there should exist an
intermediate regime (the stringy fluid)  of hot QCD that is characterized by approximate chiral spin symmetry and where degrees of freedom are chirally
symmetric quarks bound into color singlet objects by the chromoelectric field.
Above this intermediate regime the color charge and the chromoelectric field
are Debye screened and one observes a transition to QGP with magnetic
confinement.}

\FullConference{%
 The 38th International Symposium on Lattice Field Theory, LATTICE2021
  26th-30th July, 2021
  Zoom/Gather@Massachusetts Institute of Technology
}

\begin{document}
\maketitle

\section{Introduction}

Before the RHIC era there was a belief 
that at some critical temperature $T_c$ there should exist a 
deconfinement phase transition from the hadron gas to the 
quark gluon plasma (QGP) with free deconfined quarks and gluons.
There is no dynamical breaking of chiral symmetry  in a system
of free quarks and consequently the spontaneously broken
chiral symmetry below $T_c$ should be restored above $T_c$.
So the critical temperature $T_c$ should be a common temperature of
both deconfinement and of chiral restoration.
At the RHIC era this view changed since both on the lattice and
experimentally  no phase transition was seen and instead
there was a smooth but rather fast crossover observed on the lattice
first in Wuppertal \cite{W}. 
The quark condensate drops from its vacuum value at $T \sim 100$ MeV
to practically zero at $T \sim 200$  MeV with a pseudocritical
temperature around $T_{pc} \sim 155$ MeV. The Polyakov loop showed
an inflection point only slightly above this temperature so after RHIC
a new lore arised that in QCD there is a fast common deconfinement - chiral
restoration  crossover from hadron gas to QGP around the pseudocritical temperature $T_{pc} \sim 155$ MeV. This picture was also confirmed
by the Bielefeld  lattice group.

However, the inflection point of the nonrenormalized Polyakov loop was
used to establish a "deconfinement temperature". If the renormalized
Polyakov loop is studied, which is physical, see Fig. 1 \cite{P}, then one clearly sees that
there is no hint of a deconfinement transition between 100 and 200 MeV,
since the deconfinement transition should be accompanied by Polyakov loop
evolution from 0 to 1. The renormalized Polyakov loop shows its
evolution from small values to 1 in a broad temperature interval with
inflection point around 300 MeV. This temperature coincides with the
temperature of the center symmetry ("deconfinement") phase transition
in a pure glue theory. Hence the evolution of the renormalized
Polyakov loop suggests that in full QCD with light quarks a dramatic
rearrangement in gluodynamics happens still around 300 MeV, which is
however strongly smeared out by dynamical light quarks. This suggests
that properties of hot QCD should be influenced by two independent
temperatures, the chiral phase transition around $ T_c \sim 130 $ MeV \cite{K}
and by center symmetry phase transition at  $ T_d \sim 300 $ MeV. There
should be no  unique pseudocritical temperature of combined chiral restoration
and deconfinement crossover.

\begin{figure}
\centering
\includegraphics[angle=0,width=0.3\linewidth]{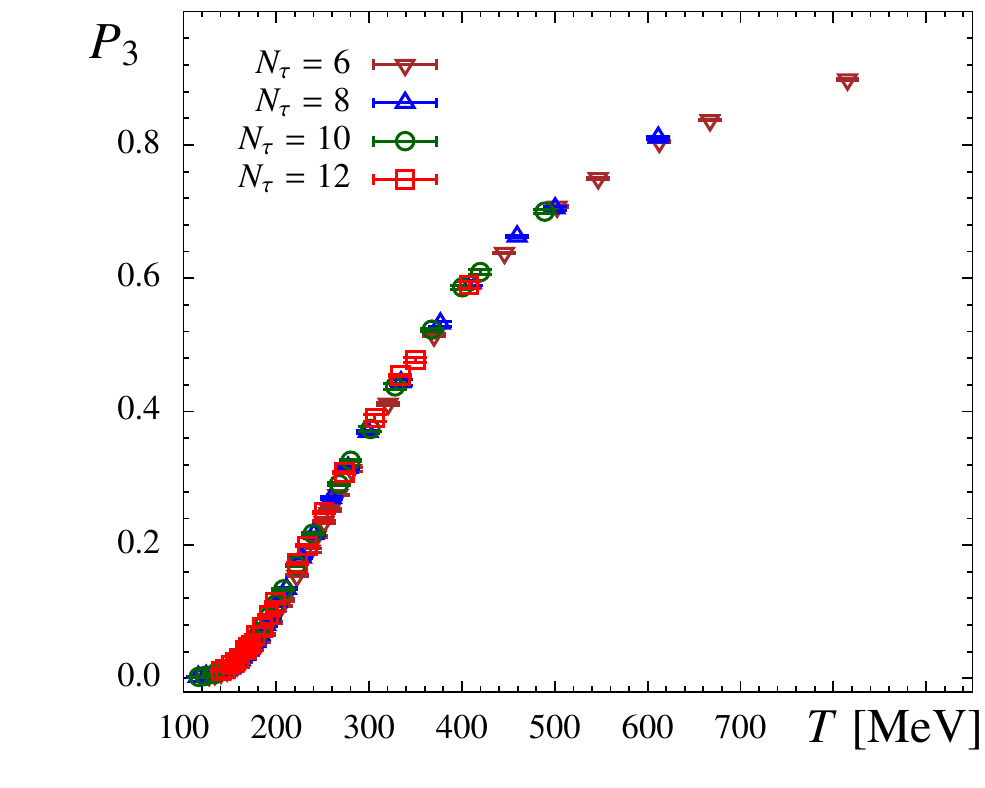}
\caption{ Renormalized Polyakov loop at physical quark masses. From
Ref. \cite{P}}
\label{St}
\end{figure}

From the Polyakov loop correlators one can extract an effective potential
between  static sources (which is however  dependent from an assumption
whether such a potential is real or complex). Such a potential below and
above $T_{pc}$ at
physical and smaller than physical quark masses is shown in Fig. 2. \cite{La}.
We clearly see essentially the same potential at both temperatures.
In both cases there is a flattening of the linear part of the potential
that is related to the string breaking. At the same time there is not
yet Debye screening of the color charge. The Debye screening is by definition
a screening of the negative Coulomb potential. Deconfinement
could be associated with the Debye screening of the color charge \cite{S}.
 Hence this potential tells that
both below and above $T_{pc}$ QCD is in the confining regime. 

\begin{figure}
\centering
\includegraphics[angle=0,width=0.3\linewidth]{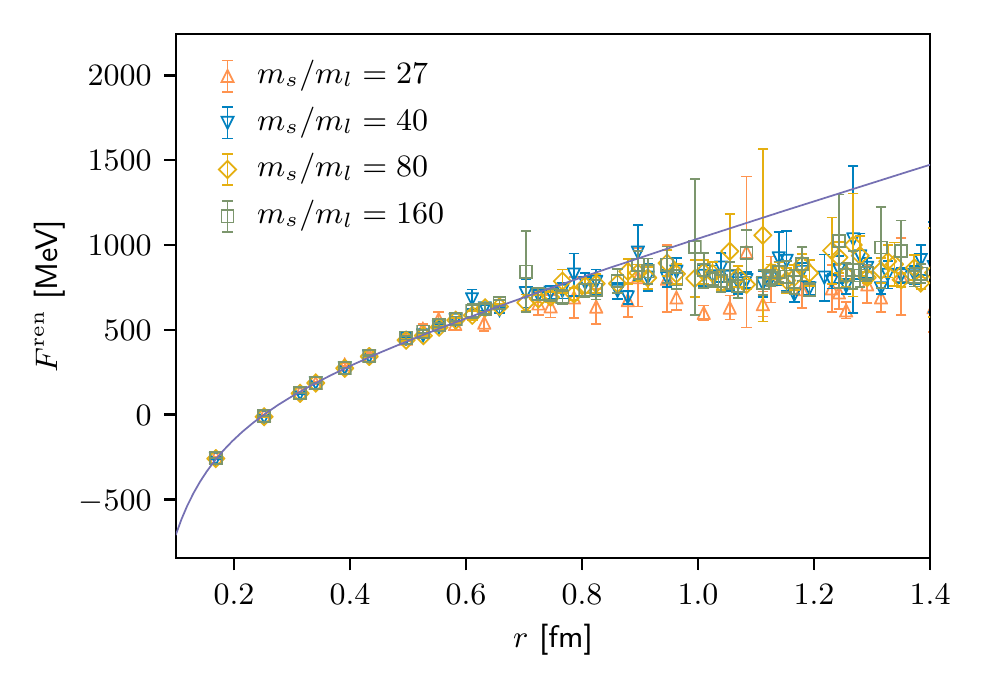}
\includegraphics[angle=0,width=0.3\linewidth]{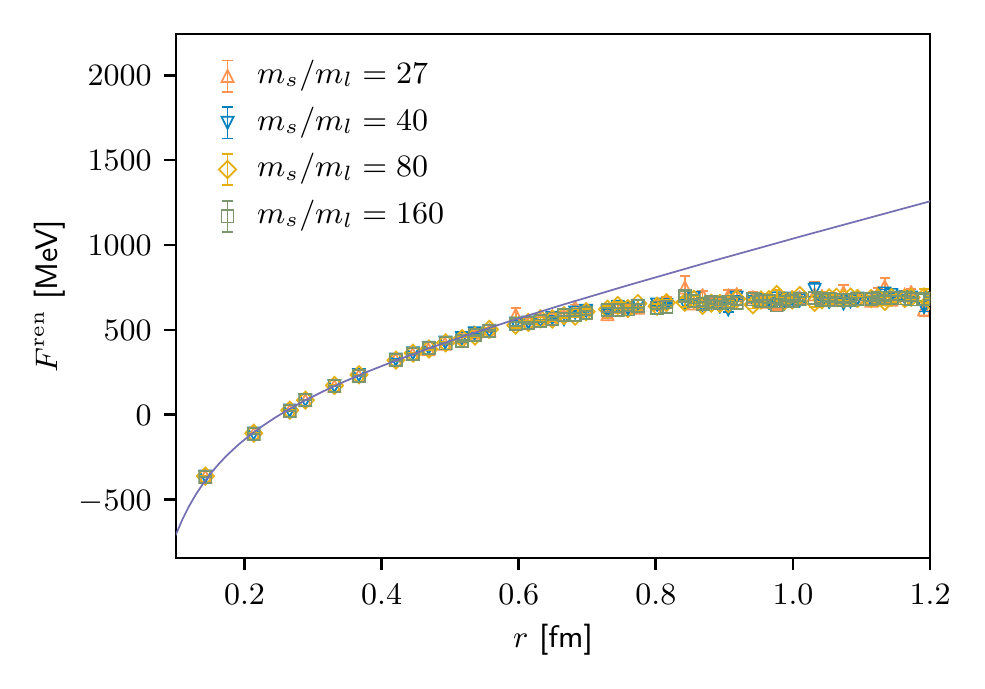}
\caption{ Effective potential between the static sources at physical
and below than physical quark masses extracted
from the correlators of renormalized Polyakov loops.
Left: $T = 141$ MeV; right: $T = 166$ MeV. From Ref. \cite{La}}
\label{Karsch}
\end{figure}

Another hint that in QCD there is no a simple fast crossover from
hadron gas to QGP is suggested by experiments. The hadron gas is a gas.
The QGP is also a gas with a large $\eta/s$. Experimentalists tell us that at RHIC and LHC
temperatures between $ \sim 150$ and $ \sim 400$ MeV they observe a
liquid with a very small $\eta/s$. Hence what is seen in experiments cannot be a QGP.

One obviously needs an objective information about degrees of freedom
above $T_{pc}$ to decide what physics and regimes take place there.
The objective and model independent information can be delivered
by symmetries that can be extracted from correlators on the lattice.

\section{Symmetries of the color charge and chromoelectric
interaction in QCD}

The  chromoelectric interaction is defined via the color charge (Lorentz-invariant)

\begin{equation}
 Q^a = \int d^3x  
\psi^\dagger(x) \frac{t^a}{2} \psi(x).
\end{equation}
\noindent
The color charge  has both $SU(N_F)_L \times SU(N_F)_R$
and $U(1)_A$ chiral symmetries which are also symmetries of the QCD Lagrangian.
However it has an additional symmetry: the chiral spin $SU(2)_{CS}$
symmetry \cite{G1,G2}.\footnote{This symmetry was reconstructed from a
large degeneracy of hadrons observed on the lattice
upon artifical truncation of the near zero modes of the Dirac operator at zero temperature \cite{D1,D2}.} 

The  $SU(2)_{CS}$  chiral spin transformations are defined as follows:

  \begin{equation}
  \psi \rightarrow  \psi^\prime = \exp \left(i  \frac{\varepsilon^n \Sigma^n}{2}\right) \psi  \; 
  \end{equation}

\noindent
with generators constructed from Euclidean gamma matrices
$\gamma_k;~~~ k=1,2,3,4$
 
 \begin{equation}
   \Sigma = \{\gamma_k,-i \gamma_5\gamma_k,\gamma_5\}.   
\end{equation}

The direct product of $SU(2)_{CS}$ and of $SU(N_F)$
can be embedded into $SU(2N_F)$ group that contains as subgroups
both $SU(N_F)_L \times SU(N_F)_R$
and $U(1)_A$ symmetries. The $SU(2N_F)$ is also a symmetry of the color 
charge.

In Minkowski space in a given reference frame the quark-gluon interaction part of the QCD Lagrangian 
can be split into temporal and spatial parts:
\begin{equation}
\overline{\psi}   \gamma^{\mu} D_{\mu} \psi = \overline{\psi}   \gamma^0 D_0  \psi 
  + \overline{\psi}   \gamma^i D_i  \psi .
\label{cl}
\end{equation}
The covariant derivative  $D_{\mu}$  includes
interaction of the quark field $\psi$ with the  gluon field $A^a_\mu$,
\begin{equation}
D_{\mu}\psi =( \partial_\mu - ig \frac{t^a  A^a_\mu}{2})\psi.
\end{equation}
The temporal term contains  interaction of the color-octet charge density
\begin{equation}
\bar \psi (x)  \gamma^0  \frac{t^a}{2} \psi(x) = \psi (x)^\dagger  \frac{t^a}{2} \psi(x)
\label{den}
\end{equation}
with the chromoelectric part of the gluonic  field. 
It is a singlet under  the $SU(2)_{CS}$  and  $SU(2N_F)$ transformations. 
The spatial part consists of a quark kinetic term and interaction with the
chromomagnetic part of the gluonic field.
It breaks  $SU(2)_{CS}$  and  $SU(2N_F)$.
Hence  interaction of quarks with the electric and magnetic components of the 
gluonic
field  can be distinguished by symmetry. 
Obviously one needs to fix a reference frame to make a discussion
of electric and magnetic components sensible.  

Emergence of approximate chiral spin and $SU(2N_F)$ symmetries in hadrons
at zero temperature upon truncation of the near-zero modes of the
Dirac operator observed in \cite{D1,D2} tells that while the magnetic interaction is located
predominantly in the near-zero modes of the Dirac operator, a confining
electric interaction is distributed among all modes of the Dirac operator.
The quark condensate of the vacuum is associated only with the near-zero
modes. Hence one can conclude from those results that confinement and
chiral symmetry breaking in QCD are not directly related phenomena.  
Given this observation it was predicted that 
above $T_{pc}$, where the chiral symmetry is restored and the near-zero
modes of the Dirac operator are suppressed by temperature effects,
there should emerge the
chiral spin  and $SU(2N_F)$ symmetries and hence QCD should
be still in the confining regime \cite{G3}, for an overview see \cite{G4}. 
In order to understand physics and degrees of freedom in QCD above
$T_{pc}$ one should study symmetry properties of  correlators
with respect to chiral spin and $SU(2N_F)$ transformations.

\section{Emergence of the chiral spin and $SU(2N_F)$ symmetries
in spatial correlators.}

In this section we shortly overview the results on symmetries of
spatial correlators obtained with JLQCD domain wall $N_F=2$
configurations at physical quark masses in Refs. \cite{R1,R2}
that have been reported at two previous lattice conferences.

In Fig. 3 we show a complete set of $J=0,1$ correlators
at different temperatures. One observes three distinct multiplets
above the chiral symmetry restoration crossover and below
roughly $T \sim 500$ MeV. The multiplet $E_1$ represents degenerate
scalar and pseudoscalar correlators. This signals effective
restoration of $U(1)_A$ symmetry that was previously observed
in Ref. \cite{JLQCD1}. The approximately degenerate
multiplets $E_2$ and $E_3$ arise from emerged approximate $SU(2)_{CS}$ and
$SU(4)$ symmetries.  These symmetries suggest that degrees
of freedom in QCD above the chiral crossover but below $T \sim 500$ MeV
are chirally symmetric quarks connected by the chromoelectric field
into color singlet objects. This regime of QCD is referred to as a stringy
fluid. At $T \sim 500$ MeV the chiral spin and $SU(4)$ symmetries disappear
and only chiral symmetries remain. It suggests that the color charge and electric field get Debye screened and one observes a smooth transition to 
QGP with quasiquarks and quasigluons being the effective degrees
of freedom. 

\begin{figure}
  \centering

  \includegraphics[width=0.49\linewidth]{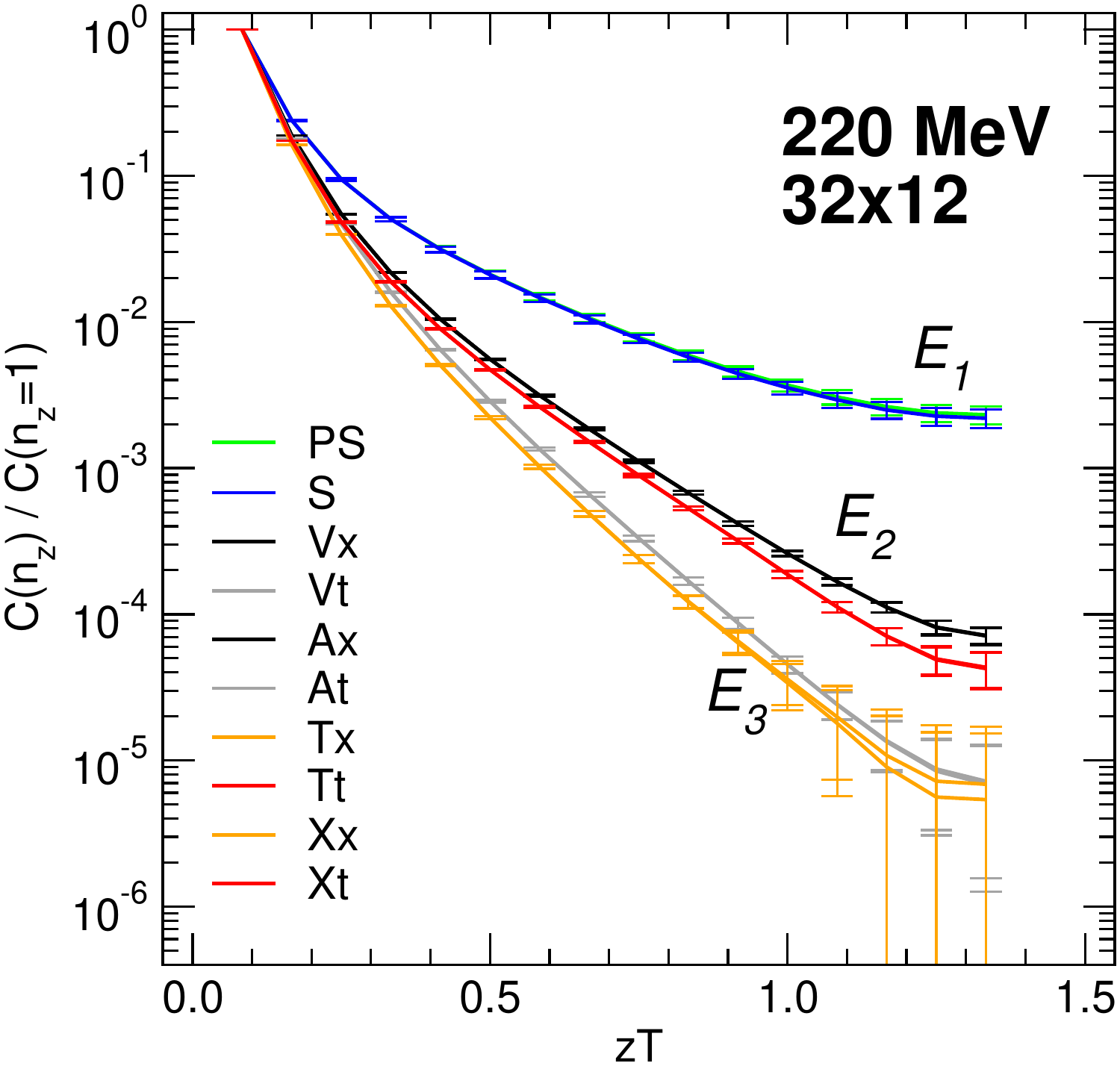}
   \includegraphics[width=0.49\linewidth]{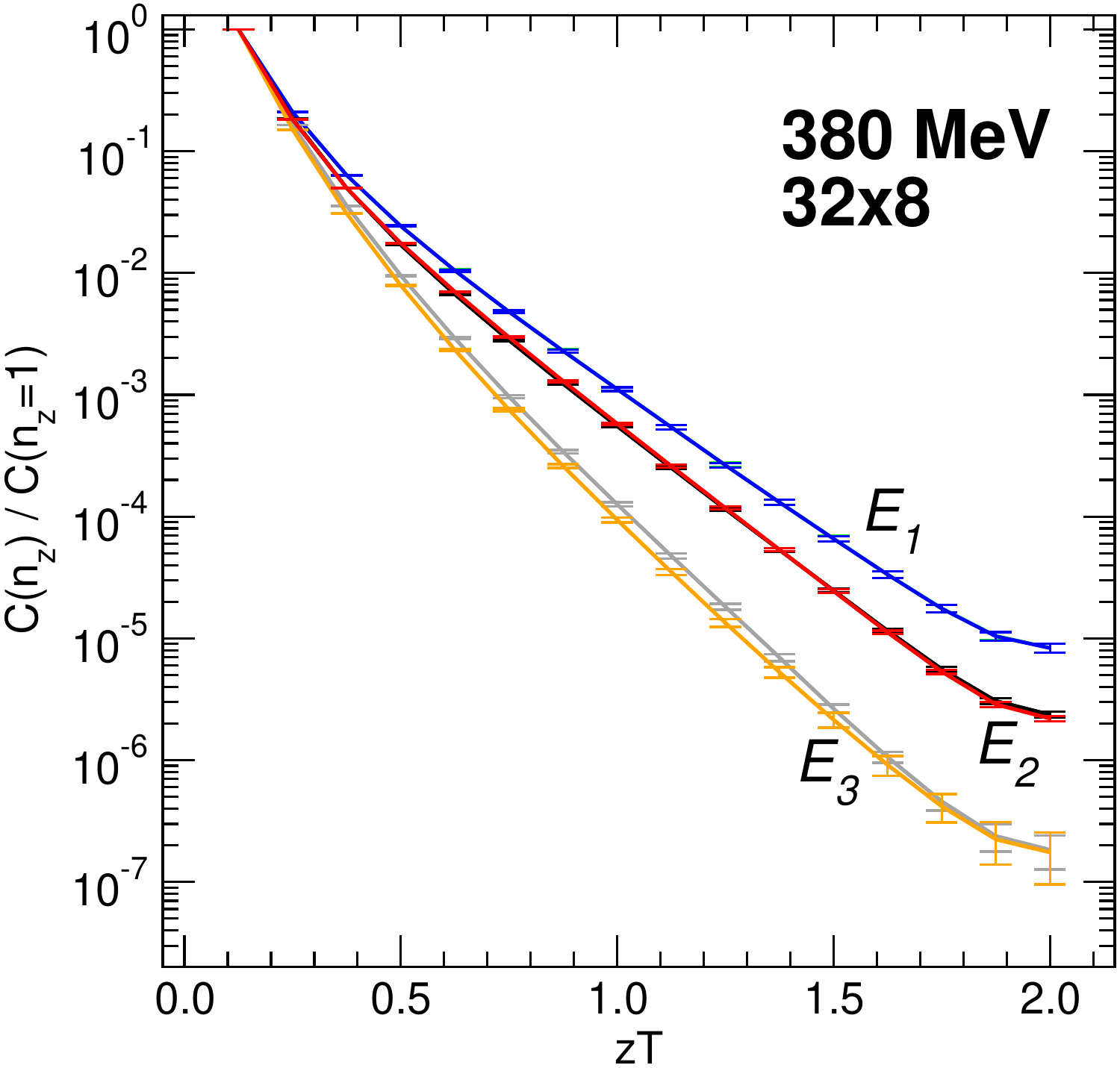}
     \includegraphics[width=0.49\linewidth]{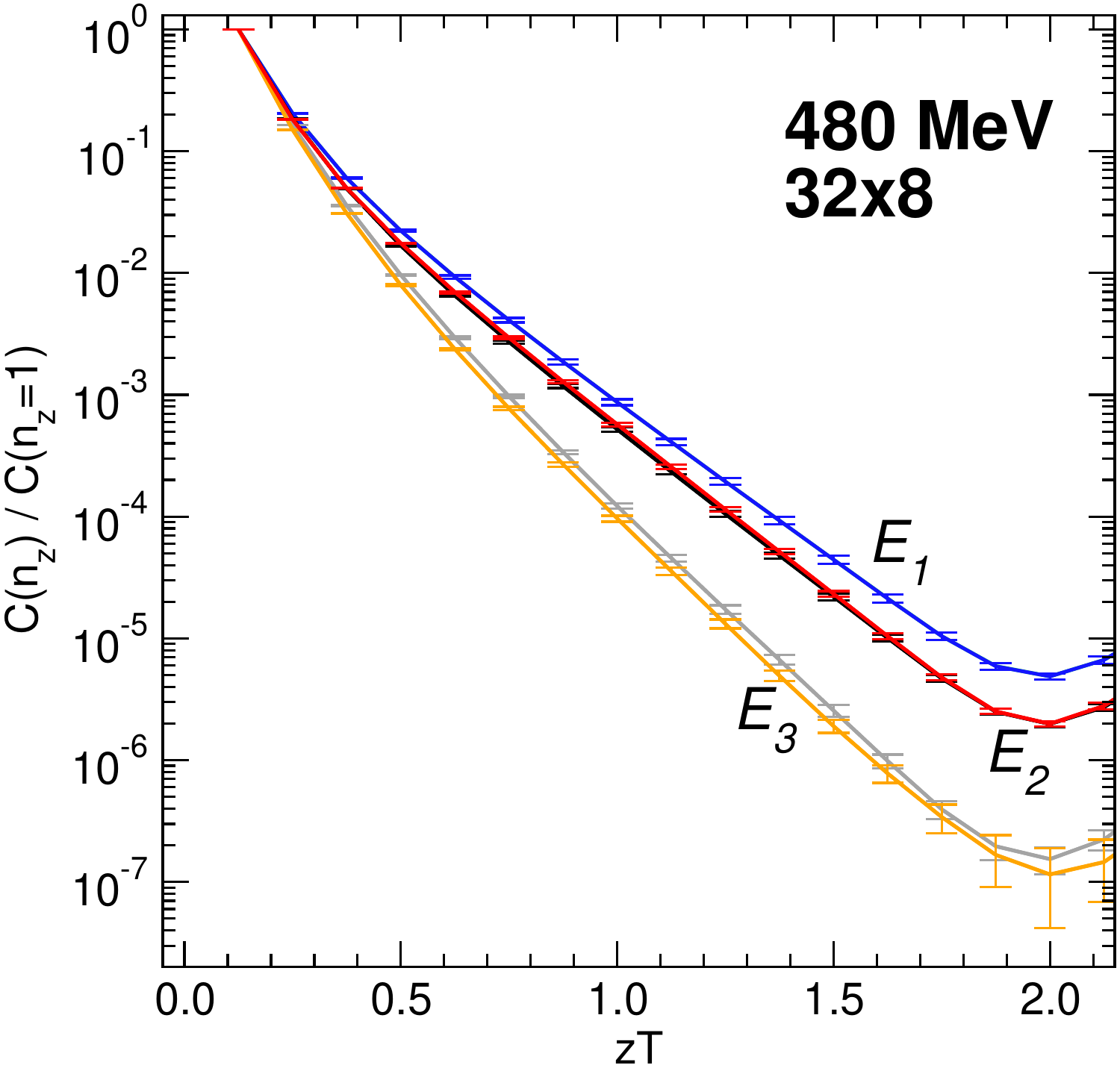}
        \includegraphics[width=0.49\linewidth]{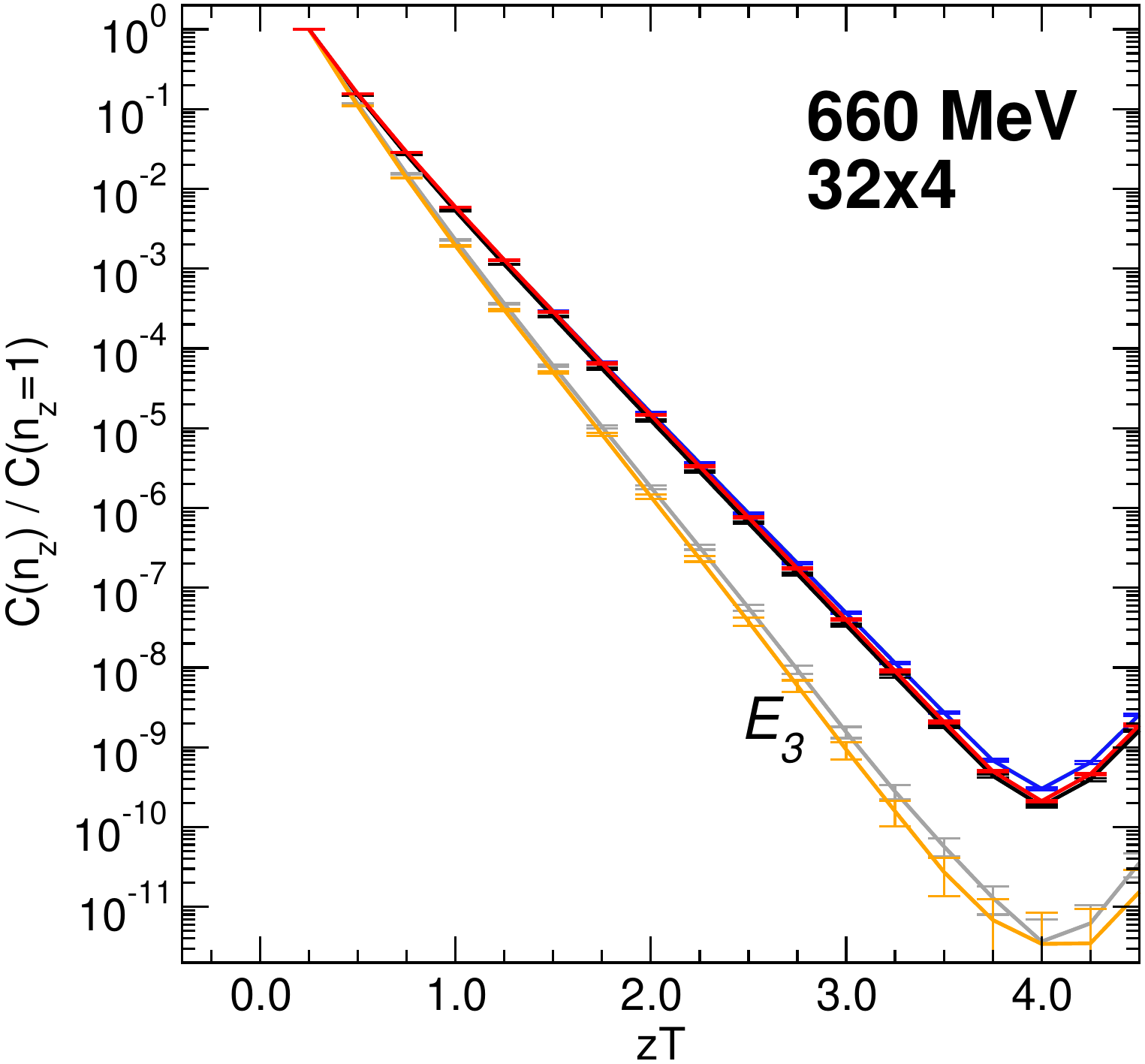}
\caption{A complete set of spatial J=0,1 correlators at different temperatures.
From Ref. \cite{R2}} 
\end{figure}

\section{Temporal correlators above $T_{pc}$}

If above the chiral crossover the QCD action is approximately
chiral spin and $SU(4)$ symmetric then it should also be observed
in temporal correlators that are connected to observable spectral
density.

The t-direction  correlators in the hadron rest
frame are

\medskip
\begin{equation}
C_\Gamma(t) = \sum\limits_{x, y, z}
<\mathcal{O}_\Gamma(x,y,z,t)
\mathcal{O}_\Gamma(\mathbf{0},0)^\dagger>,
\nonumber
\end{equation}

\noindent
where $\mathcal{O}_\Gamma = \bar q \Gamma \frac{\vec{\tau}}{2} q$
are all possible $J=0$ and $J=1$ local operators.
The transformation properties of the $J=1$ operators
with respect to $U(1)_A$, $SU(2)_L \times SU(2)_R$, $SU(2)_{CS}$
and $SU(4)$ are shown in Fig. 4 \cite{G2}. Emergence of these
symmetries should be seen as a degeneracy of the corresponding
correlators.
\begin{figure}
  \centering
  \includegraphics[width=0.49\linewidth]{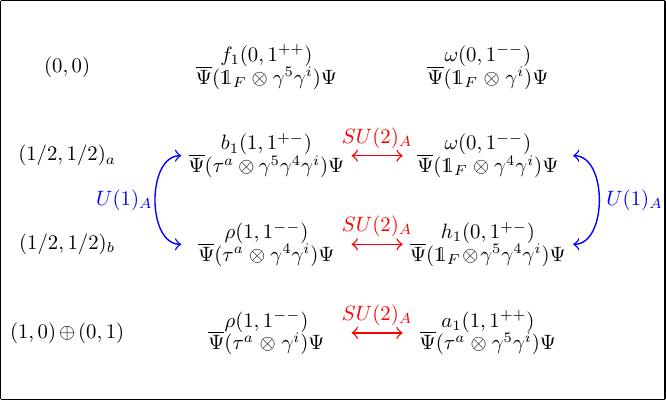}
  \includegraphics[width=0.49\linewidth]{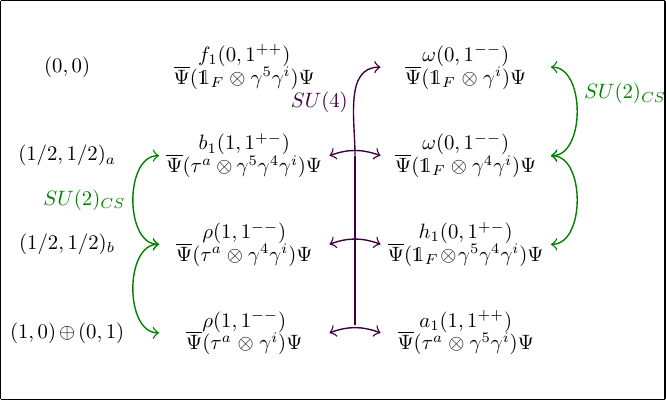}
  \label{transformations}
\caption{Transformations between interpolating vector operators, $i=1,2,3$.
  The left columns indicate the chiral representation for each operator.
  Red and blue arrows connect operators that transform into
  each other under $SU(2)_L \times SU(2)_R$ and  $U(1)_A$, respectively.
  Green arrows connect operators that form triplets of $SU(2)_{CS}, k=4$.
  The $f_1$ and $a_1$ operators are the $SU(2)_{CS}, k=4$ -- singlets.
  Purple arrows show the 15-plet of $SU(4)$. The $f_1$ operator is
  a $SU(4)$-singlet. From Ref. \cite{G2}}
\end{figure}

These correlators with the domain wall Dirac operator at physical quark
masses with $N_F=2$ QCD (JLQCD configurations)  normalized at $n_t=1$
calculated on  $48^3 \times 12$ lattices at $220$ MeV \cite{R3}
are shown in Fig. 5. Note that one needs a sufficiently large
size $N_t$ along the time direction to observe a real  evolution
of the temporal correlation functions. 
E.g. with $N_t=2$
by construction only chiral symmetries can be obtained since everything is fixed
by the free quark fields at source and sink.
Consequently at the moment
we are limited only to the temperature $T=220$ MeV. 

On the left side of Fig.~5 we show the correlators calculated with
 free, noninteracting quarks. Dynamics
 of free quarks are governed by the Dirac equation and only chiral
 symmetries exist. Indeed  only
 degeneracies due to $U(1)_A$ and $SU(2)_L \times SU(2)_R$ symmetries
 are seen in meson correlators calculated for free quark gas. This pattern
 reflects correlators at very high temperatures
 since due to the asymptotic freedom at a very high $T$ the quark-gluon
 interactions can be neglected.
 
 On the right side of Fig.~5 we show the correlators calculated
 in full QCD at $T=220$ MeV. The pattern in full QCD is qualitatively
 different as compared to the free quark gas. In full QCD one
 clearly sees multiplets of all symmetries under discussion:
 $U(1)_A$, $SU(2)_L \times SU(2)_R$, $SU(2)_{CS}$ and $SU(4)$.
 This confirms results obtained with the spatial correlators.

\begin{figure}
\centering
  \includegraphics[width=0.49\linewidth]{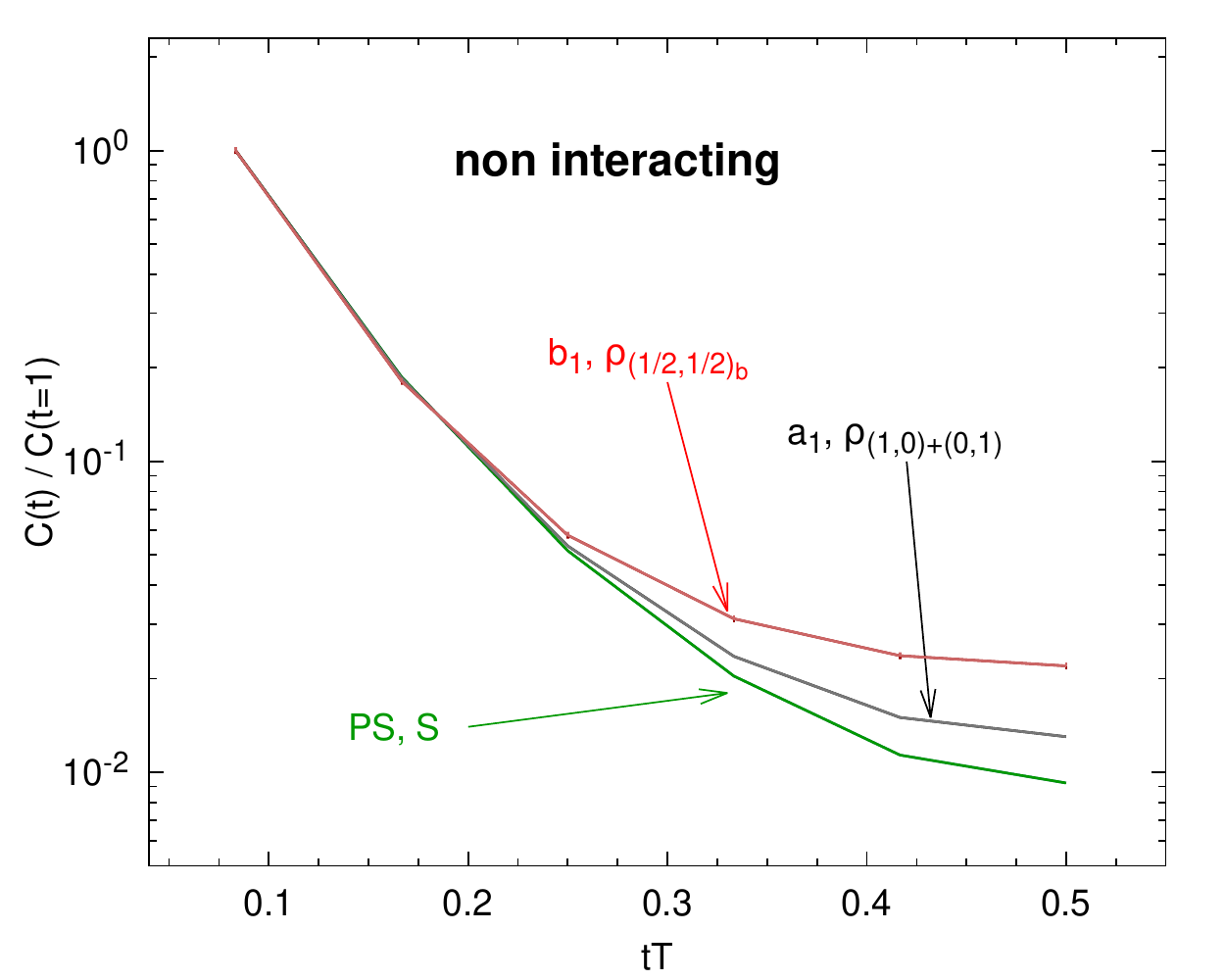}
  \includegraphics[width=0.49\linewidth]{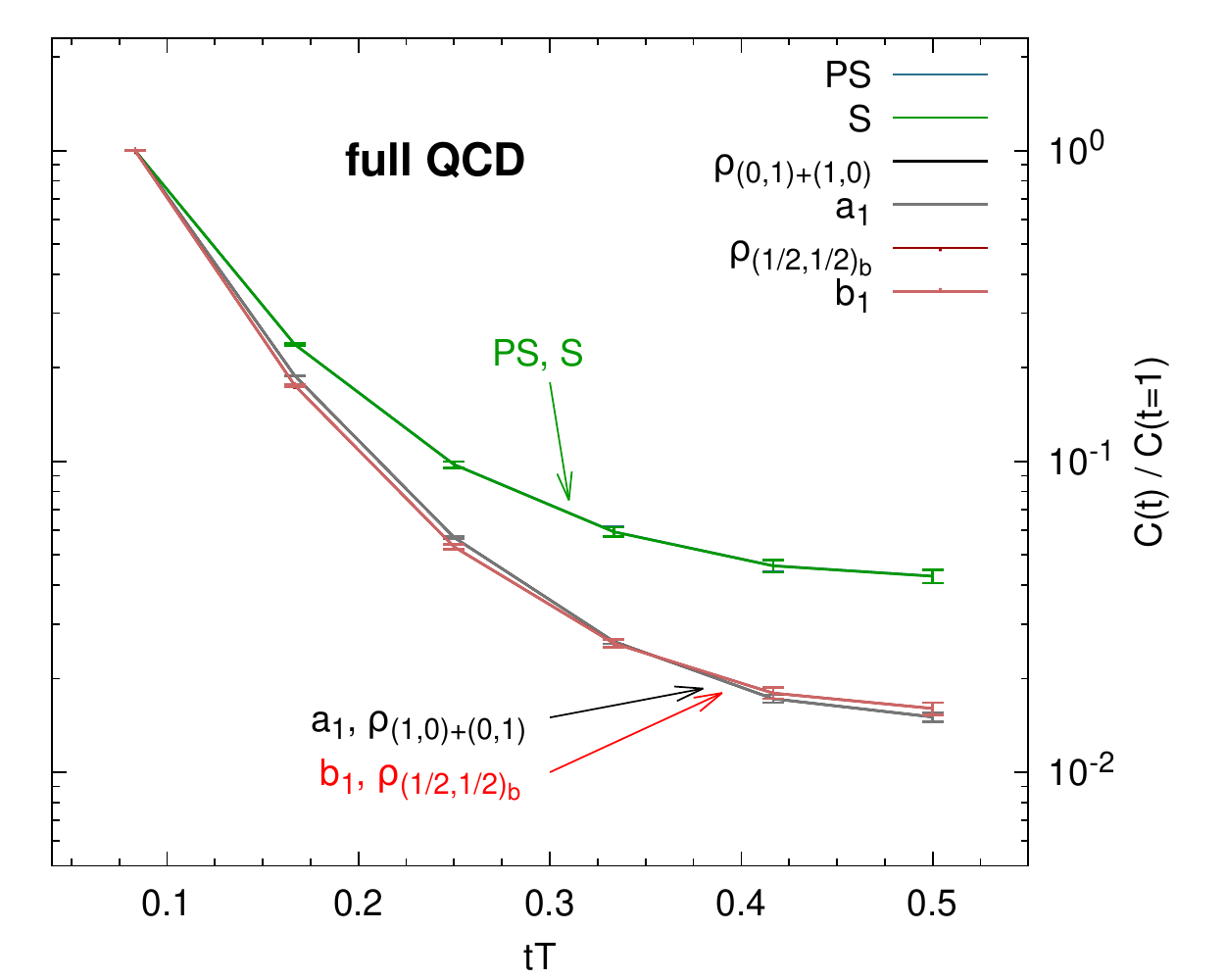}
 \caption{Temporal correlation functions for $48^3\times 12$ lattices.
          The l.h.s. shows correlators calculated with free noninteracting quarks on the
          same lattice, and features a
          symmetry pattern expected from chiral symmetry.
          The r.h.s. presents full QCD data at a temperature of $T=220$ MeV,
          which shows multiplets of all $U(1)_A$, $SU(2)_L \times SU(2)_R$,
	  $SU(2)_{CS}$ and $SU(4)$ groups. From Ref. \cite{R3}}
  \label{tcorrs}
\end{figure}

\section{Conclusions}

Our results on symmetries of correlators of $N_F=2$
QCD at physical quarks masses allow to distinguish
three different regimes of QCD at nonzero temperatures.

\begin{figure}
\centering
\includegraphics[width=0.49\linewidth]{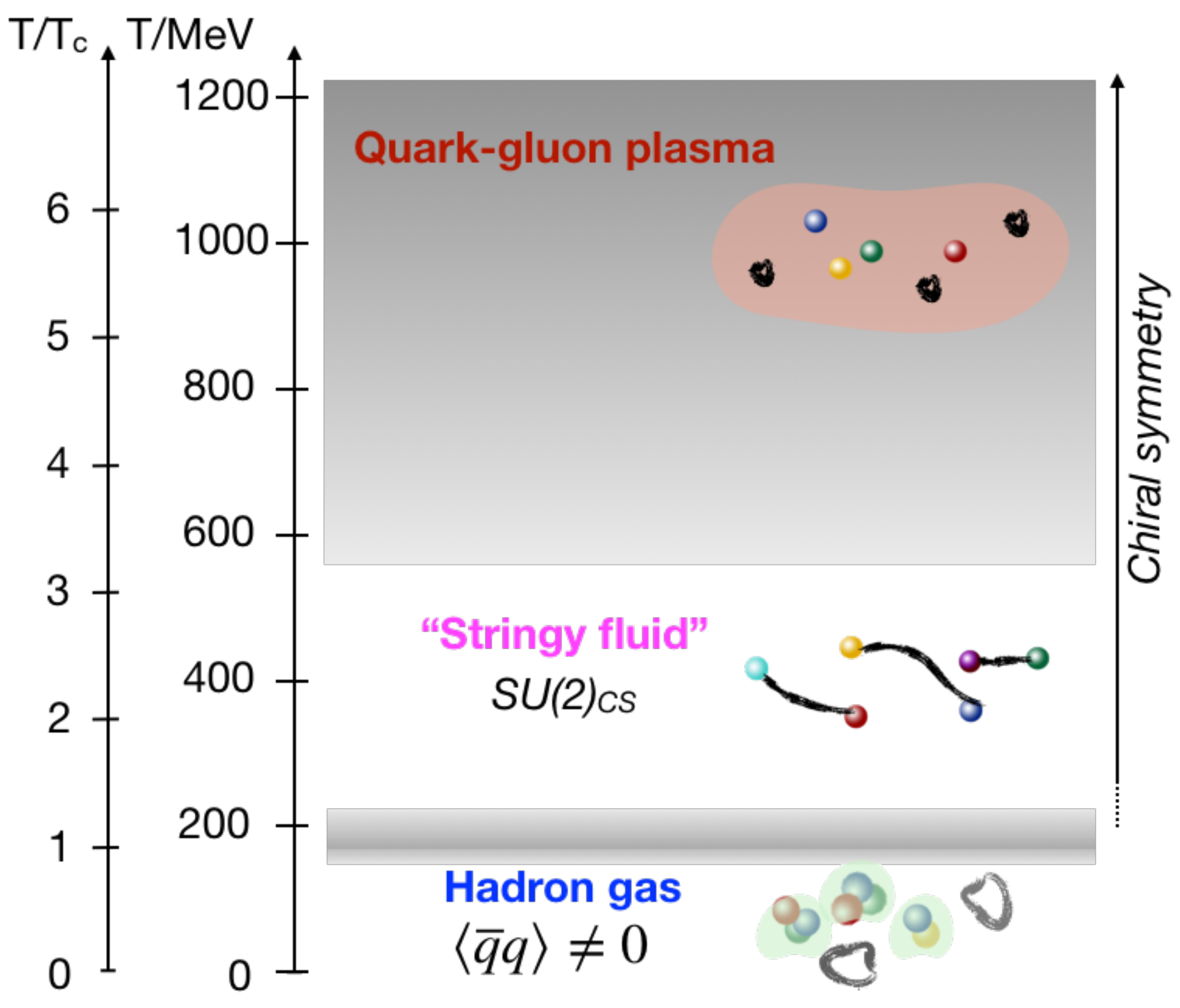}
\caption{Three regimes of QCD. From Ref. \cite{R2}}.
\end{figure}

At temperatures $0 - T_{pc}$  we have a  hadron gas consisting of confined mesons
with broken chiral symmetry. From the hadron gas there is a smooth
chiral symmetry restoration crossover and one arrives at the 
stringy fluid 
that is characterized by chiral, $SU(2)_{CS}$ and $SU(4)$ symmetries. The
electric confinement still persists and degrees of freedom are chirally
symmetric quarks bound by the electric field into color singlet compounds.
Around $T \sim 3 T_{pc}$ the color charge gets Debye screened,
the $SU(2)_{CS}$ and $SU(4)$ symmetries disappear (only chiral symmetries
remain) and one
observes a very smooth transition to QGP with quasiquarks and quasigluons
being effective degrees of freedom.

\end{document}